\documentclass[english,conference]{IEEEtran}
\usepackage[T1]{fontenc}
\usepackage[latin9]{inputenc}
\usepackage{mathtools}
\usepackage{amsmath}
\usepackage{xcolor}
\usepackage{amsthm}
\usepackage{amssymb}
\usepackage{floatrow}
\usepackage{hyperref}
\makeatletter
\theoremstyle{plain}
\newtheorem{thm}{\protect\theoremname}
\theoremstyle{definition}
\newtheorem{defn}[thm]{\protect\definitionname}
\theoremstyle{definition}

\theoremstyle{plain}
\newtheorem{lem}[thm]{\protect\lemmaname}

\usepackage{stmaryrd}

\usepackage{tikz,pgfplots,pgf}
\pgfplotsset{compat=1.17}
\usetikzlibrary{calc,fadings,decorations.pathreplacing}
\usetikzlibrary{3d}
\usetikzlibrary{arrows.meta}
\usepackage{ifthen}
\usepackage{overpic}
\usepackage{latexsym,amsmath,xcolor,multicol,booktabs,calligra}
\usepackage{graphicx}
\usepackage{wrapfig}

\usetikzlibrary{arrows,decorations.pathmorphing,decorations.footprints,decorations.pathreplacing,fadings,calc,trees,mindmap,shadows,decorations.text,patterns,positioning,shapes,matrix,fit}

\usepackage{color}
\definecolor{bitcolor}{rgb}{1,0.84314,0}
\definecolor{checkcolor}{rgb}{0.52941,0.80784,1}

\makeatother

\usepackage{babel}
\providecommand{\definitionname}{Definition}
\providecommand{\lemmaname}{Lemma}
\providecommand{\problemname}{Problem}
\providecommand{\theoremname}{Theorem}

\IEEEoverridecommandlockouts

\begin{document}

\title{Belief Propagation with Quantum Messages\\
for Symmetric Classical-Quantum Channels}

\author{S. Brandsen, Avijit Mandal, and Henry D. Pfister\thanks{\hrule\vspace{1mm}The authors are with Duke University in the Department of Electrical and Computer Engineering. Please direct e-mail to \url{henry.pfister@duke.edu}.}}

\maketitle
\global\long\def\cB{\mathcal{B}}

\global\long\def\cD{\mathcal{D}}

\global\long\def\cH{\mathcal{H}}

\global\long\def\cQ{\mathcal{Q}}

\global\long\def\cX{\mathcal{X}}

\global\long\def\cY{\mathcal{Y}}

\global\long\def\bra#1{\left\langle #1\right|}

\global\long\def\ket#1{\left|#1\right\rangle }

\global\long\def\braket#1#2{\left\langle #1\middle|#2\right\rangle }

\global\long\def\sketbra#1{\ket{#1}\mkern-4mu\bra{#1}}
\global\long\def\ketbra#1#2{\ket{#1}\mkern-4mu\bra{#2}}

\global\long\def\cnop{\boxast}

\global\long\def\vnop{\varoast}

\newcommand{\Tr}{\mathrm{Tr}}
\newcommand{\px}{\sigma_{\mathrm{x}}}
\newcommand{\py}{\sigma_{\mathrm{y}}}
\newcommand{\pz}{\sigma_{\mathrm{z}}}

\newcommand{\Henry}[1]{\textcolor{blue}{HP: #1}}
\newcommand{\Sarah}[1]{\textcolor{purple}{SB: #1}}
\newcommand{\Avijit}[1]{\textcolor{green}{AM: #1}}

\begin{abstract}
Belief propagation (BP) is a classical algorithm that approximates the marginal distribution associated with a factor graph by passing messages between adjacent nodes in the graph.
It gained popularity in the 1990's as a powerful decoding algorithm for LDPC codes.
In 2016, Renes introduced a belief propagation with quantum messages (BPQM) and described how it could be used to decode classical codes defined by tree factor graphs that are sent over the classical-quantum pure-state channel.
In this work, we propose an extension of BPQM to general binary-input symmetric classical-quantum (BSCQ) channels based on the implementation of a symmetric ``paired measurement''.
While this new paired-measurement BPQM (PMBPQM) approach is suboptimal in general, it provides a concrete BPQM decoder that can be implemented with local operations.

Finally, we demonstrate that density evolution can be used to analyze the performance of PMBPQM on tree factor graphs.
As an application, we compute noise thresholds for some LDPC codes with BPQM decoding for a class of BSCQ channels.

\end{abstract}

\vspace{-1mm}
\section{Introduction}
\vspace{-1mm}

Low-density parity-check (LDPC) codes and iterative decoding were introduced by Gallager in 1960~\cite{Gallager-60} but they did not attract widespread interest until the introduction of Turbo codes~\cite{Berrou-icc93} and the rediscovery of LDPC codes~\cite{Mackay-it99} four decades later.

Belief propagation (BP), in its general form, was introduced by Pearl in 1982~\cite{Pearl-aaai82} as an efficient algorithm to exactly compute marginals for tree-structured probability models. BP works by passing messages between neighboring nodes and it was subsequently shown that BP includes both Gallager's iterative decoding and turbo decoding as special cases~\cite{McEliece-jsac98,Kschischang-jsac98}.

Starting in the 1990s, there has been a growing interest in generalizing concepts from classical coding to the case of classical quantum (CQ) channels. Holevo, Schumacher, and Westmoreland identified the maximum rate of classical information transfer over a CQ channel~\cite{PhysRevA.56.131}.
Subsequent works described code constructions and decoding strategies that achieve this optimal rate~\cite{Wilde2013, Lloyd2012}.
At the same time, advances in photonic communication underscored the need for developing low-complexity decoding protocols~\cite{PhysRevA.105.032446, PhysRevLett.118.040801, da_Silva_2013, PhysRevLett.106.240502}.
In particular, there is a significant gap between the information rate achievable by a receiver with individual pulse-by-pulse detection and the rate possible with optimal joint quantum receiver (i.e., measurement of the full output system)  if the mean number of photons per received optical pulse
is smaller than one.

A key question is whether generalizations of BP can also be used to efficiently decode codes transmitted over CQ channels.
Just as direct computation of the marginal probability distribution is computationally infeasible for large factor graphs in the classical case, it is experimentally infeasible to naively implement the optimal (Helstrom) measurement for decoding a code defined by a large factor graph.
This is because it would require many quantum operations involving all the qubits.
The idea of generalizing BP to decode a classical binary code transmitted over a CQ channel was introduced by Renes~\cite{Renes-njp17} and it offers an alternative to experimentally infeasible collective measurements.
The described approach is restricted to the pure-state channel (PSC) and codes whose Tanner graphs are trees.
It is described based on a channel-combining perspective that is also adopted in this work.  Following~\cite{Rengaswamy-isit20b}, we will refer to this general decoding method as belief-propagation with quantum messages (BPQM). In~\cite{Rengaswamy-isit20b}, simulation results are presented for a simple 5-bit code (whose Tanner graph is a tree) and compared to a classical decoding approach. For the 5-bit code, it is observed in~\cite{Rengaswamy-isit20b} (and proved in~\cite{Rengaswamy-npjqi21}) that bit-optimal decoding is actually block optimal in this context. However, this version of BPQM had exponential complexity due to the need for controlled unitary operations which grow with the size of the tree.

Piveteau and Renes significantly advanced the understanding of BPQM in~\cite{Piveteau-arxiv21}, where they prove that BPQM is simultaneously optimal both for bit-error probability and block-error probability for binary linear codes with tree factor graphs on the PSC. They additionally reduce the overall decoding complexity from exponential to quadratic by introducing the idea of using a quantum reliability register for each qubit message.

In this work, we introduce a paired-measurement BPQM (PMBPQM) protocol which is the first extension of BPQM to general symmetric binary-input CQ (BSCQ) channels. This extension is based on a lemma that shows any BSCQ channel can be approximated by an orthogonal mixture of BSCQ channels that output a single qubit. In some ways, this is similar to the fact that any symmetric binary-input classical channel can be represented as a stochastic mixture of binary symmetric channels (BSCs)~\cite[p.~182]{RU-2008}. For classical bit-flip channels, it is easy to verify that PMBPQM simplifies to classical BP and, for the PSC, we also show that PMBPQM simplifies to the BPQM defined in~\cite{Renes-njp17}.

For more general BSCQ channels, we also have an example that shows \emph{any} approach involving binary-outcome measurements will be suboptimal relative to the collective Helstrom success probability. However, by comparing the performance of PMBPQM to the Helstrom measurement for a variety of factor graphs with up to 13 qubits, we observe that PMBPQM is near optimal for the chosen instances.  One interesting open question is ``What is the worst-case gap between PMBPQM and the collective Helstrom measurement?''.

We also analyze the performance of PMBPQM for large factor graphs by deriving its density evolution equations.
Density evolution, which was formalized in~\cite{Richardson-it01}, is an asymptotic analysis method that can be used to find noise thresholds for successful BP decoding (in terms of bit-error rate\footnote{Add details}) of long LDPC codes sent through symmetric channels.
For LDPC codes on CQ channels with optimal Helstrom decoding, there are no currently known methods for computing channel noise thresholds.
We demonstrate that PMBPQM lends itself to asymptotic analysis and characterize the region of channel parameters for a qubit BSCQ channel such that asymptotically reliable decoding is achievable via PMBPQM.

\vspace{-2mm}
\section{Notation}
\vspace{-1mm}

We define the set of natural numbers by $\mathbb{N}=\left\{ 1,2,\ldots\right\} $ and use the shorthand $[m]\coloneqq\left\{ 1,\ldots,m\right\} $ for $m\in\mathbb{N}$. We denote the $n$-dimensional complex Hilbert space by $\mathcal{H}_{n}$ and write the $i$-th element of the standard basis of $\mathcal{H}_n$ as $\ket i$ for $i \in \left\{ 0,1,.\ldots,n-1\right\} $. A pure quantum state, $\ket{\psi}\in\mathcal{H}_n$, is an $n$-dimensional complex vector with unit length. 
The Hermitian transpose of $\ket{\phi}\in\mathcal{H}_{n}$ is denoted either by $\ket{\phi}^{\dag}$ or $\bra{\phi}$. The inner product between $\ket{\psi}$ and $\ket{\phi}$ is denoted by $\braket{\phi}{\psi}\coloneqq\ket{\phi}^{\dag}\ket{\psi}$ and all pure states satisfy $\braket{\psi}{\psi}= 1$. We abuse notation and define $\ket{\theta} \coloneqq \cos(\frac{\theta}{2}) \ket{0} + \sin(\frac{\theta}{2}) \ket{1}$ by its argument $\theta$. The Pauli matrices are denoted by \[\px \coloneqq \begin{bmatrix}0 & 1 \\ 1& 0 \end{bmatrix}, \py \coloneqq \begin{bmatrix}0 & -i \\ i& 0 \end{bmatrix}, \pz \coloneqq \begin{bmatrix} 1& 0 \\ 0 & -1 \end{bmatrix} \]
and the Hadamard matrix by $H \coloneqq (\px+\pz)/\sqrt{2}$.

 A stochastic mixture of pure states is called a mixed state. Consider a random pure state defined by $\left\{ p_{i},\ket{\psi_{i}}\right\} _{i\in[m]}$ which takes value $\ket{\psi_{i}}$ with probability $p_{i}$. The associated mixed state is represented by the density matrix $\rho=\sum_{i=0}^{m-1}p_{i}\ket{\psi_{i}}\!\!\bra{\psi_{i}}$,
which is a positive semidefinite matrix with unit trace.
Let $\mathcal{D}(\mathcal{H}_{n})$ denote the set of density matrices (i.e., positive semi-definite $n\times n$ complex matrices with unit trace).
When the value of $n$ is not important, we will use $\cH$ to denote the Hilbert space and $\cD(\cH)$ to represent the set of density matrices.

Finally, we denote with $\mathcal{B}(\mathcal{H}_{n})$ the set of positive semi-definite $n \times n$ complex matrices with bounded (not necessarily unit) trace.
A quantum measurement on an $n$-dimensional quantum system is then represented by $\hat{M} = \{M_{j}\}|_{j=1}^{m}$ where each element $M_{j} \in \mathcal{B}(\mathcal{H}_{n})$ and $\sum_{j=1}^{m} M_{j} = \mathbb{I}_{n}$. In the case of projective measurements where $M_{j} M_{i} = \delta_{i, j} M_{i}$, we denote the measurement by $\hat{\Pi} = \{\Pi_{j}\}|_{j=1}^{m}$.

\vspace{-1mm}
\subsection{Classical Quantum Channels}
\vspace{-1mm}

\begin{defn}
\label{CQdef}
A BSCQ channel, $W\colon\left\{ 0,1\right\} \to\cD(\cH_{n})$, maps classical input $z \in \{0, 1\}$ to the density matrix output $W(z) \in \cD(\cH_{n})$ and obeys the symmetry constraint $W(1)=UW(0)U$ for a unitary matrix $U$ satisfying $U^{2} = \mathbb{I}$. If $n=2$, then the output lives in a 2-dimensional Hilbert space and we call this a \emph{qubit channel}.
\end{defn}

\begin{defn}
A minimum-error measurement $\hat{M} = \{M_{j}\}|_{j=1}^{m}$ for a given set of candidate states $\{\rho_{j}\}|_{j=1}^{m}$ with corresponding probabilities $p_{j} = \text{Pr}(\rho = \rho_{j})$ is a measurement that maximizes the probability of correct detection,
\vspace{-1mm}
\begin{align*}
    \sum_{j=1}^{m} p_{j} \Tr[M_{j} \rho_{j}].
\end{align*}
\end{defn}

\begin{defn}
The Helstrom measurement is a minimum-error measurement for a given binary state set $\{\rho_{+}, \rho_{-} \}$ with corresponding probabilities $\{p, 1-p \}$. The Helstrom measurement 
\begin{align*}
\hat{\Pi}_{\text{H}}(\rho_{+}, \rho_{-}, p) = \left\{\hat{\Pi}_{+}(\rho_{+}, \rho_{-}, p), \mathbb{I} - \hat{\Pi}_{+}(\rho_{+}, \rho_{-}, p)\right\}
\end{align*}
is defined by projectors onto the positive and negative eigenspaces of the matrix $p \rho_{+} - (1-p) \rho_{-}$~\cite{Helstrom-jsp69}. Namely,
\begin{align*}
\label{eq:helstrom_j}
\Pi_{+}(\rho_{+}, \rho_{-}, p) \coloneqq \sum_{\ket{v} \in \mathcal{V}_{+}(\rho_{+}, \rho_{-}, p)} \ket{v}\!\!\bra{v},
\end{align*}
where, for $M = \left( p \rho_{+} - (1-p) \rho_{-} \right)$, we define
\begin{align*}
    \mathcal{V}_{+}(\rho_{+}, \rho_{-}, p) \coloneqq \left\{ \ket{v} \!\in\! \cH \, \big\vert \braket{v}{v}\!=\!1, \exists \lambda\geq 0, M \ket{v} = \lambda \ket{v} \right \}.
\end{align*}
We denote the corresponding success probability by
 \begin{align*}
     P_{\text{H}}(\rho_{+}, \rho_{-}, p) &\coloneqq p \text{Tr}\left[ \hat{\Pi}_{+}(\rho_{+}, \rho_{-}, p) \rho_{+} \right] \\
     & + (1-p) \text{Tr}\left[ \left(\mathbb{I} - \hat{\Pi}_{+}(\rho_{+}, \rho_{-}, p)\right)\rho_{-} \right].
 \end{align*}
\end{defn}

\section{Paired-Measurement BPQM}

\subsection{Representation of Symmetric CQ Channels}

\begin{lem}
\label{symmetricCQ}
Any qubit BSCQ channel is unitarily equivalent to $W \colon \left\{ 0,1\right\} \to\cD(\mathcal{H}_{2})$ satisfying $W(z)=\px^{z}\rho(\theta, p)\px^{z}$ with
\[
\rho(\theta,p) \coloneqq (1-p) H\sketbra{\theta} H^{\dag} + \frac{p}{2} \mathbb{I}.
\]
\end{lem}

\begin{IEEEproof}
By definition, a qubit BSCQ satisfies $W(z) = U^{z} \rho_{0} U^{z}$ for a unitary $U$ satisfying $U^{2} = \mathbb{I}$ and a qubit density matrix $\rho_{0}$.
Since unitary matrices are unitarily diagonalizable (i.e., $U=V\Lambda V^\dag$) and $U^2 = \mathbb{I}$, we observe that
\begin{align*}
\tilde{W}(z) &\coloneqq H V^\dag W(z) V H^\dag \\
&= H \Lambda^z ( H^\dag H ) (V^\dag \rho_0 V )(H^\dag H) \Lambda^z H^\dag \\
&= (H \Lambda^z H^\dag) \tilde{\rho}_0 (H \Lambda^z H^\dag),
\end{align*}
where $\tilde{\rho}_0 =H V^\dag \rho_0 V H^\dag$ and  $\Lambda$ is diagonal with $\pm 1$ entries on the diagonal.
If $\Lambda = \pm \mathbb{I}$, then $\tilde{W}(0)=\tilde{W}(1)$ and the  channel is trivial.
In all other cases, no generality is lost if we choose $\Lambda = \pz$ and this gives $\tilde{W}(z) = \px^z \tilde{\rho}_0 \px^z$.

Now, we will show that $\px^{z} \tilde{\rho}_{0} \px^{z}$ is unitarily equivalent to $\px^{z} \rho(\theta, p) \px^{z}$. Since any qubit state can be written as a depolarized pure state, there are parameters $(p', \theta', \phi')$ such that 
\begin{align*}
    \tilde{\rho}_{0} = (1-p') H\sketbra{\theta', \phi'}H^{\dag} + p' \frac{\mathbb{I}}{2},
\end{align*}
where $\ket{\theta, \phi}\coloneqq \cos(\frac{\theta}{2}) \ket{0} + e^{i \phi} \sin(\frac{\theta}{2})\ket{1}$. Let $U_{1}$ be the unitary satisfying $U_1 \Big( H\ket{\pm \theta', \phi'}\Big) = H\ket{\pm \theta}$ with $|\braket{\theta', \phi'}{ - \theta', \phi'}| = |\braket{\theta}{ - \theta}|$ (i.e., $U_{1} H$ is equivalent to rotation about the $\mathrm{z}$-axis of the Bloch sphere by $-\phi'$).

Then, we have 
  \begin{align*}
 U_{1} & \px^{z} \tilde{\rho}_{0} \px^{z} U_{1}^{\dag} \\ &= (1-p') U_{1} \px^{z} \Big( H \sketbra{\theta', \phi'}H^{\dag}\Big) \px^{z} U_{1}^{\dag} + p' \frac{\mathbb{I}}{2} \\
 &= (1-p') U_{1} \Big( H \sketbra{(-1)^{z} \theta', \phi'} H^{\dag}\Big) U_{1}^{\dag} + p' \frac{\mathbb{I}}{2} \\
 &= (1-p') H\sketbra{(-1)^{z} \theta} H^{\dag} + p' \frac{\mathbb{I}}{2},
 \end{align*}
 which is equivalent to $\px^{z} \rho(\theta, p) \px^{z}$.
\end{IEEEproof}

\vspace{1.5mm}

We now introduce a ``paired measurement'' that processes the outputs of an arbitrary BSCQ channel into an orthogonal combination of qubit BSCQ channels.
This illustrates the symmetry of collective Helstrom measurement.
It is also one part of PMBPQM, which uses  bit-node and check-node combining in concert with sequential applications of paired measurement. 

\begin{defn}
For a BSCQ channel $W(z) = U^{z} \rho U^{z}$ with $\rho \in \mathcal{D}(\mathcal{H}_{2n})$, let $\{\ket{v_{j}} \}|_{j=1}^{n}$ be $n$ orthogonal eigenvectors of $W(0) - W(1)$ with non-negative eigenvalues $\{\lambda_j\}|_{j=1}^n$ satisfying $(W(0)-W(1))U \ket{v_j} = - \lambda_j U \ket{v_j} $ and $U \ket{v_j} \perp \ket{v_j}$.
The existence of such a set is established in Lemma~\ref{lem:helstorm_paired} and the \emph{paired measurement} for $W$ is defined to be
\begin{align*}
    \hat{\Pi}_{W} \coloneqq \Big\{ \sketbra{v_{j}} +  U \sketbra{v_{j}} U  \Big\}\Big|_{j=1}^{n}.
\end{align*}
\end{defn}

\begin{lem} \label{lem:helstorm_paired} Consider the task of distinguishing between equiprobable outputs of a BSCQ channel $\{W(0), W(1)\}$.

Then, the optimal Helstrom measurement $\hat{\Pi}_{H}$ for this channel is equivalent to first implementing the paired measurement for $W$ and then using its outcome, $j$, to select the second measurement,
$\hat{\Pi}_{W}(j) =  \big\{\sketbra{v_{j}},  U \sketbra{v_{j}}U \big\}$.
\end{lem}
\begin{IEEEproof}
Let $M = \rho - U \rho U$ and $\ket{v_{j}}$ be an eigenvector of $M$ with eigenvalue $\lambda_{j}$. Then, $U\ket{v_{j}}$ is an eigenvector with eigenvalue $-\lambda_{j}$ because
\begin{align*}
    M (U \ket{v_{j}}) &= (\rho U  - U \rho) \ket{v_{j}}
    = U (U \rho U  - \rho) \ket{v_{j}} \\ &= - U M \ket{v_{j}} = - \lambda_{j} U \ket{v_j}.
\end{align*}

The Helstrom measurement is determined by the positive and negative eigenspaces of $M$. Thus, one can choose an orthonormal basis $\{\ket{v_j}\}|_{j=1}^{n}$ for the $n$ largest eigenvalues $\{\lambda_{j}\}|_{j=1}^{n}$ such that $\bra{v_{j}} U \ket{v_{j}} = 0$ and the Helstrom measurement is given by 
\begin{align*}
    \hat{\Pi}_{H} = \Bigg\{\sum_{j=1}^{n} \sketbra{v_{j}}, \sum_{j=1}^{n} U \sketbra{v_{j}} U  \Bigg\}.
\end{align*}

For the case where 0 is an eigenvalue of $M$, we still need to justify that such a choice is possible.
Let $S$ be the zero eigenspace of $M$ and consider the natural projective measurement $\{ \Pi_S, \Pi_{S^\perp} \}$ designed to distinguish between $S$ and $S^\perp$.
Conditioned on the outcome associated with $S$, the trace distance between post-measurement density matrices is 0 because the conditional error rate is $\frac{1}{2}$ (i.e., $S$ only contains the parts of $\rho$ and $U\rho U$ that are indistinguishable).
This implies that $\rho_S \coloneqq \Pi_S \rho \Pi_S = \Pi_S U \rho U \Pi_S$.
Also, since $-0=0$, the first argument shows that $S$ is mapped to itself by $U$.
It follows that $\Pi_S U = U \Pi_S$ and $\rho_S = U \rho_S U$.
Moreover, these identities also imply that $U \rho_S = \rho_S U$.
Thus, $U$ and $\rho_S$ are simultaneously diagonalizable.

Let the unitary $T$ diagonalize $U$ and $\rho_S$ so that $\Lambda = T^H U T$ and $\Upsilon = T^H \rho_S T$ are diagonal.
Then, the diagonal entries of $\Lambda$ associated with eigenvectors supported on $S$ are $\pm 1$ because $U^2 = \mathbb{I}$.
This implies that $U \rho_S U = (T \Lambda T^H) (T \Upsilon T^H) (T \Lambda T^H) = T \Upsilon T^H$ is independent of $\Lambda$.
Hence, any choice of $\Lambda$ gives rise to an equivalent channel defined by $\rho$ and $\tilde{U} = \Pi_{S^\perp} U \Pi_{S^\perp} + \Pi_S T \Lambda T^H \Pi_S$
satisfying $U \rho U = \tilde{U} \rho \tilde{U}$.
It follows that we can choose $\Lambda$ to have exactly half $1$ and $-1$ diagonal entries.
To put the eigenvectors of the zero eigenspace into the stated form, assume $\ket{u}$ and $\ket{u'}$ are eigenvectors of $\Pi_{S}T\Lambda T^{H} \Pi_{S}$ with eigenvalues $+1$ and $-1$ respectively. Then, $\ket{u}$ and $\ket{u'}$ are in $S$ and $\ket{v_j} = \frac{1}{\sqrt{2}}(\ket{u} + \ket{u'})$ satisfies $\tilde{U} \ket{v_j} \perp \ket{v_j}$.

The overall statement follows from noting that the Helstrom measurement can also be implemented by a rank-one projective measurement $\hat{\Pi} = \big\{ \ket{v_{j}}\bra{v_{j}}, U\ket{v_{j}}\bra{v_{j}}U  \big\}\big|_{j=1}^{n}$ which is equivalent to the paired measurement followed by measuring the post-measurement state with subspace with $\hat{\Pi}_{W}(j)$. 
\end{IEEEproof}

\begin{lem}
For a BSCQ channel $W\colon \{0, 1\} \rightarrow \mathcal{D}(\mathcal{H}_{2n})$ with equiprobable inputs, the channel $W$ followed by paired measurement $\hat{\Pi}_{W}$ is equivalent to a BSCQ which defines the classical mixture of symmetric qubit channels given by
\[
\tilde{W}(z) = \sum_{j=1}^{n}p_{j}\Big(\px^{z}\rho_{j}\px^{z}\otimes\sketbra{j} \Big).
\]
From this, we see that the $j$-th paired outcome has probability $p_j = \Tr\big[\big(\sketbra{v_{j}} + U \sketbra{v_{j}} U\big) \rho \big]$ and results in a post-measurement density matrix equivalent to 
\[ \rho_j = \frac{1}{p_j} \begin{pmatrix}
    \bra{v_{j}} \rho \ket{v_{j}} & \bra{v_{j}} U \rho \ket{v_{j}} \\
    \bra{v_{j}} \rho U \ket{v_{j}} & \bra{v_{j}} U \rho U \ket{v_{j}}
    \end{pmatrix}. \]
\end{lem}

\begin{IEEEproof} 
The probability that implementing $\hat{\Pi}_{W}$ on $W(z)$ yields outcome $j$ (corresponding to $\hat{\Pi}_{\text{W}}(j) = \ket{v_{j}}\!\!\bra{v_{j}} + U^{x} \ket{v_{j}}\!\!\bra{v_{j}} U^{x}$) is: 
\begin{align*}
    P(j|z) &= \Tr\Big[\Big(\sketbra{v_{j}} + U \sketbra{v_{j}} U\Big) U^{z}\rho U^{z} \Big] \\ &= \Tr\Big[U^{z}\Big(\sketbra{v_{j}} + U \sketbra{v_{j}}U\Big) U^{z}\rho \Big] \\ & = \Tr\Big[\Big(\sketbra{v_{j}} + U \sketbra{v_{j}} U\Big) \rho \Big].
\end{align*}
As this is independent of $z$, we denote it simply as $P(j)$. The corresponding post-measurement state may be written as
\begin{align*}
    \tilde{\rho}(j, z) & \coloneqq \frac{1}{P(j)} \hat{\Pi}_{\text{W}}(j) U^{z} \rho U^{z} \hat{\Pi}_{\text{W}}(j)
\end{align*}
It follows that $\tilde{\rho}(j, z)$ is effectively a two-dimensional state with all nonzero components spanned by the basis $\{\ket{v_{j}}, U\ket{v_{j}}\}$. Then an equivalent qubit state is given by
\begin{align*}
    \tilde{\rho}(j, z) &= \frac{1}{P(j)} \px^{z} \begin{pmatrix}
    \bra{v_{j}} \rho \ket{v_{j}} & \bra{v_{j}} U \rho \ket{v_{j}} \\
    \bra{v_{j}} \rho U \ket{v_{j}} & \bra{v_{j}} U \rho U \ket{v_{j}}
    \end{pmatrix} \px^{z}.
\end{align*}
 For simplicity, we denote $\tilde{\rho}(j,0)$ as $\rho_{j}$ and denote with $\{\lambda_{k}^{(j,z)}\}$ the set of eigenvectors for $\rho_{j}$ where $\big\{ \ket{\lambda_{k}(j,z)} \big\}$ is the corresponding set of eigenvectors. Then, conditioned on input $z$, the final state after paired measurement is $\sum_{j} p_{j} \px^{z} \rho_{j} \px^{z} \otimes \ket{j}\!\!\bra{j}$.
\end{IEEEproof}

 \subsection{Channel Combining for Bit and Check Nodes}
 
 The following definitions of bit-node and check-node combining can be found in~\cite{Renes-it18} and are the natural generalizations of classical definitions for LDPC codes. 
Since these operations preserve channel symmetry, the paired-measurement can also be applied to the combined channels.
 
\begin{defn}
For CQ channels $W,W'$, the bit-node and check-node channel combining operations are defined by
\begin{align*}
[W\vnop W'](z) & \coloneqq W(z)\otimes W'(z)\\
[W\cnop W'](z) & \coloneqq\frac{1}{2}\sum_{z'\in\{0,1\}}W(z\oplus z')\otimes W'(z').
\end{align*}
\end{defn}
\begin{lem}
If $W,W'$ are symmetric CQ channels, then $W\vnop W'$ and $W\cnop W'$ are symmetric CQ channels.
\end{lem}
\begin{IEEEproof}
By assumption, there exists $U,U'$ such that $W(1)=UW(0)U$ and $W'(1)=U'W'(0)U'$. Thus, for bit-node combining, we have
\begin{align*}
(U\otimes & U')  [W\vnop W'](0)(U\otimes U') \\ &=(U\otimes U')\big(W(0)\otimes W'(0)\big)(U\otimes U') =[W\vnop W'](1).
\end{align*}
This implies that $W\vnop W'$ has the unitary symmetry operator $T=U\otimes U'$. Likewise, for check-node combining, we have
\begin{align*}
(U\otimes I) & \big([W\cnop W'](0)\big)(U\otimes I) \\ &=\frac{1}{2}(U\otimes I)(\sum_{z'\in\{0,1\}}W(z')\otimes W'(z'))(U\otimes I) \\ 
 & =\frac{1}{2}\sum_{z'\in\{0,1\}}W(1\oplus z')\otimes W'(z') \\
 &=[W\cnop W'](1).
\end{align*}
 This implies that the channel $W\cnop W'$ has the unitary symmetry operator $T=U\otimes I$.
\end{IEEEproof}

\subsection{Optimality of PMBPQM for Pure-State Channels (PSC)}
We show that implementing paired-measurements before bit- and check-node combining is equivalent on PSCs to the BPQM method outlined in~\cite{Renes-njp17, Rengaswamy-isit20b}. Given that a coherent implementation of this BPQM method has been proven to be optimal for PSCs~\cite{Piveteau-arxiv21}, it follows that paired-measurement BPQM is likewise optimal for PSCs. First, we define the canonical PSC $W_{\theta}$.

\begin{defn}
The canonical pure state channel (PSC), $W_{\theta}$, maps binary input $z \in \{0, 1\}$ to $W_{\theta}(z) = \px^{z} (H\sketbra{\theta} H^{\dag}) \px^{z}$.
\end{defn}

Now, we show that using paired-measurement BPQM after check combining is equivalent to coherent BPQM check node combining~\cite{Renes-njp17} via applying $\text{CNOT}_{1\rightarrow 2}$.
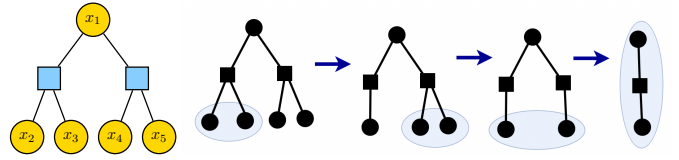
\begin{figure}[t]
\centering
\vspace{3mm}
\begin{minipage}{.25\textwidth}
 \begin{center}
  \scalebox{0.6}{%
  \begin{tikzpicture}%
      [scale=0.65,var/.style={fill=bitcolor,draw,circle,thick,minimum size=4mm},%
      factor/.style={fill=checkcolor,draw,rectangle,thick,minimum size=5mm},%
      weight/.style={font=\small}]

      \node (x1) [var] at (0,0) {$x_1$};
      \node (f2) [factor] at (1.5,-2) {};
      \node (f1) [factor] at (-1.5,-2) {};
          \node (x2) [var] at (-2.25,-4) {$x_2$};
          \node (x3) [var] at (-0.75,-4) {$x_3$};
          \node (x4) [var] at (0.75,-4) {$x_4$};
          \node (x5) [var] at (2.25,-4) {$x_5$};
          \path[draw,thick] (x1) -- node[left] {} (f1);
          \path[draw,thick] (x1) -- node[right] {} (f2);
          \path[draw,thick] (f1) -- node {} (x2);
          \path[draw,thick] (f1) -- node {} (x3);
          \path[draw,thick] (f2) -- node {} (x4);
          \path[draw,thick] (f2) -- node {} (x5);
    \end{tikzpicture}}
    \end{center}
\end{minipage}%
\begin{minipage}{.75\textwidth}
  \centering
  \includegraphics[width=0.98\linewidth]{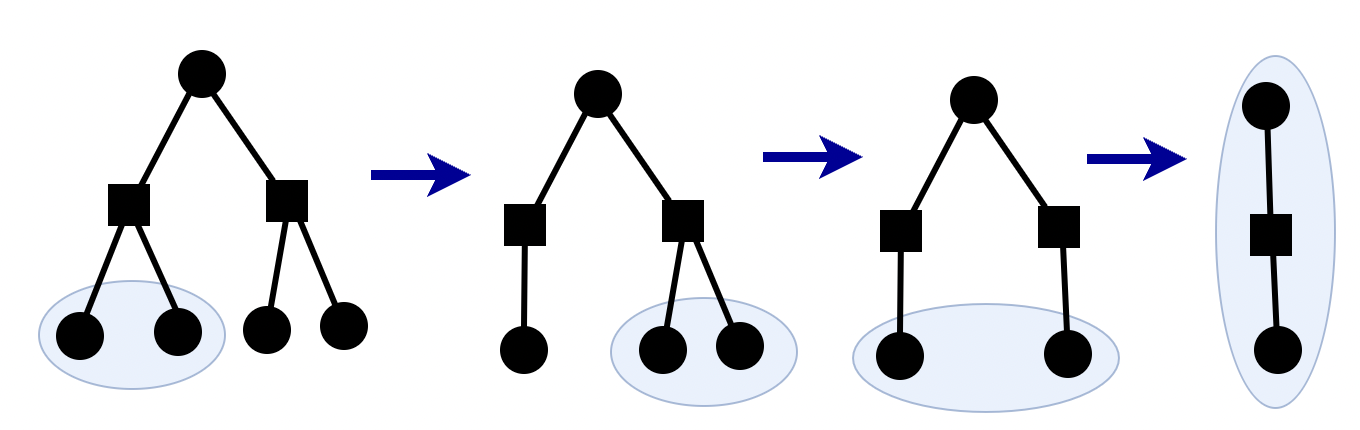}
\end{minipage}  
  
  \caption{\label{fig:fg5} The left side shows the five-qubit factor-graph. The right side depicts paired-measurement BPQM for the five-qubit factor-graph, where each stage of paired-measurement BPQM merges two qubits into one new qubit.}
\end{figure}

\begin{figure*}[t]
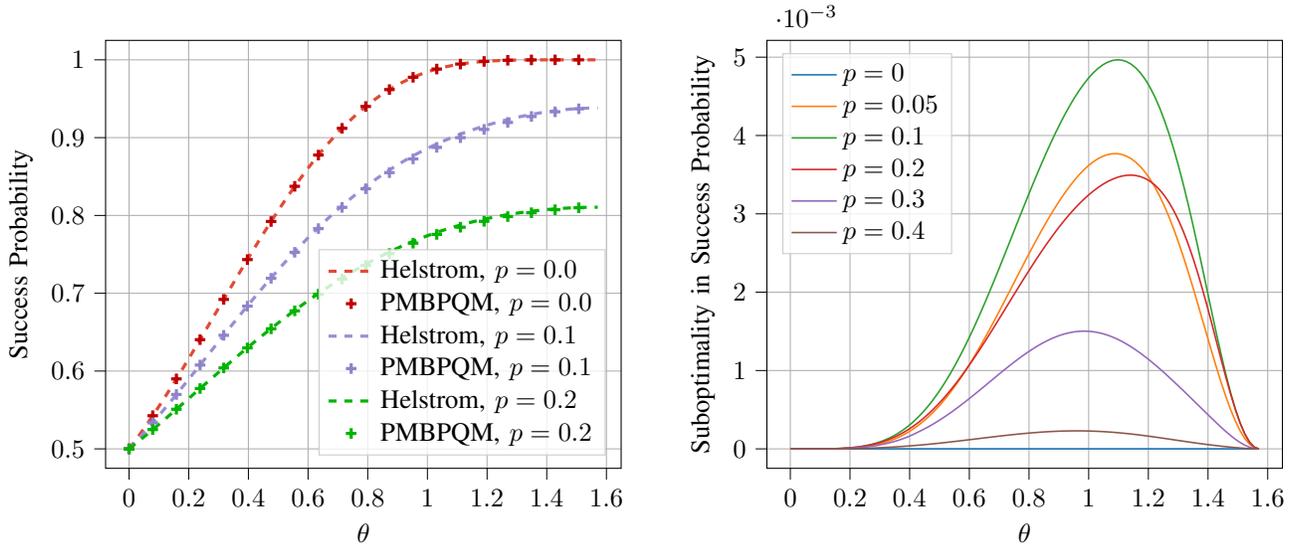

\begin{center}
  \input{density2ba.tex}
\hspace{3mm}
\input{reldiff.tex}
\end{center}
\vspace{-3mm}
\caption{\label{fig:hel_pm_reldiff}The left axis depicts success probability for both the Helstrom and paired-measurement BPQM approach as a function of $\theta$ for $p\in\{0, 0.1, 0.2\}$. The right axis shows the relative difference in success probability between the Helstrom and paired-measurement BPQM approach (i.e. $(P_{\text{helstrom} - P_{\text{BPQM}}})/{P_{\text{helstrom}}}$) as a function of $\theta$ for $p\in\{0, 0.05, 0.1, 0.2, 0.3\}$.}
\end{figure*}

\begin{lem}
Implementing the paired-measurement to distinguish between $[W_{\theta} \cnop W_{\theta'}](0)$ and $[W_{\theta} \cnop W_{\theta'}](1)$ is unitarily equivalent to implementing 
\begin{align*}
    \text{CNOT}_{1 \rightarrow 2} & [W_{\theta} \cnop W_{\theta'}](z)  \text{CNOT}_{1 \rightarrow 2} \\ &= \sum_{j \in \{0, 1\} } p_{j} \sketbra{(-1)^{z} \theta_{j}^{\cnop}} \otimes \sketbra{j},
\end{align*}
where $p_{0} = \frac{1}{2}(1+\cos(\theta) \cos(\theta'))$, $p_{1} = 1- p_{0}$, $\theta_{0}^{\cnop} = \cos^{-1} \left( \frac{\cos(\theta) + \cos(\theta')}{1 +  \cos(\theta) \cos(\theta')} \right)$, and $\theta_{1}^{\cnop} = \cos^{-1} \left( \frac{\cos(\theta) - \cos(\theta')}{1 - \cos(\theta) \cos(\theta')} \right)$.
\end{lem}
\begin{IEEEproof}
The difference matrix is given by
\begin{align*}
[W_{\theta} & \cnop W_{\theta'}](0) - [W_{\theta} \cnop W_{\theta'}](1) \\ &= \frac{1}{2} \sin (\theta) \sin (\theta') 
\begin{pmatrix}
  1 & 0 & 0 & 0 \\
  0 & -1 & 0 & 0 \\
  0 & 0 & -1 & 0 \\
  0 & 0 & 0 & 1 \\
\end{pmatrix}.
\end{align*}
The optimal choice of paired measurement is defined by
    $\ket{\tilde{v}_{0}} = \frac{1}{\sqrt{2}} (1,0,0,1)$ and  $\ket{\tilde{v}_{1}} = \frac{1}{\sqrt{2}} (-1,0,0,1)$ 
with respective eigenvalues of $\pm \sin (\theta) \sin (\theta')/2 $. The probability of obtaining outcome $\Pi_{0} = \sketbra{\tilde{v}_{0}} + U \sketbra{\tilde{v}_{0}} U $ is $p_{0} = \frac{1}{2}(1 + \cos(\theta)\cos(\theta'))$ for symmetry operator $U = \px \otimes \mathbb{I}$.
The corresponding post-measurement state for outcome $j \in \{0, 1\}$ is given by
\begin{align*}
    \tilde{\rho}(j, z) &= \px^{z} H \sketbra{- \theta_{j}^{\cnop}} H^{\dag} \px^{z}. \vspace{-5mm} \tag*{\IEEEQEDhere}
\end{align*}
\end{IEEEproof}

Using~\cite{Renes-njp17}, we can similarly demonstrate unitary equivalence of paired-measurement for bit-node combining to the coherent BPQM operation.
First, we define
\begin{align*}
 U_{\vnop}(\theta, \theta') &:= \begin{pmatrix}
   a_{+} & 0 & 0 & a_{-} \\
   a_{-} & 0 & 0 & -a_{+} \\
   0 & b_{+} & b_{-} & 0 \\
   0 & b_{-} & -b_{+} & 0
 \end{pmatrix},\\
 a_{\pm} &:= \frac{1}{\sqrt{2}} \Bigg( \frac{\cos\big(\frac{\theta - \theta'}{2}\big) \pm \cos\big(\frac{\theta + \theta'}{2}\big)}{\sqrt{1+ \cos(\theta)\cos(\theta')}} \Bigg), \\ b_{\pm} &:= \frac{1}{\sqrt{2}} \Bigg( \frac{\sin\big(\frac{\theta + \theta'}{2}\big) \mp \sin\big(\frac{\theta - \theta'}{2}\big)}{\sqrt{1 - \cos(\theta)\cos(\theta')}} \Bigg).
\end{align*}

\begin{lem}
Implementing the paired-measurement to distinguish between $[W_{\theta} \vnop W_{\theta'}](0)$ and $[W_{\theta} \vnop W_{\theta'}](1)$ is unitarily equivalent to
\begin{align*}
   U_{\vnop}(\theta, \theta') [W_{\theta} \vnop W_{\theta'}](z)U_{\vnop}(\theta, \theta')^{\dag} &:= \sketbra{(-1)^{z} \theta^{\vnop}},
\end{align*}
where $\theta^{\vnop} = \cos^{-1} \left( \cos(\theta)\cos(\theta') \right)$.
\end{lem}

\begin{IEEEproof}
The paired measurement is based on the paired eigenvectors of the difference matrix 
\begin{align*}
    &[W_{\theta} \vnop W_{\theta'}](0) - [W_{\theta} \vnop W_{\theta'}](1)  \\
    &= \begin{pmatrix}
       -h & f(\theta, \theta') & f(\theta', \theta) & 0 \\
 f(\theta, \theta') & -g & 0 & f(\theta', \theta) \\
 f(\theta', \theta) & 0 & g & f(\theta, \theta') \\
 0 & f(\theta', \theta) & f(\theta, \theta') & h \\
    \end{pmatrix}
\end{align*}
where $f(\theta, \theta') = -\frac{1}{2}\sin(\theta)\cos(\theta')$, $g =  \frac{1}{2} ( \sin(\theta') - \sin(\theta)) $, and $h = \frac{1}{2}(\sin(\theta) + \sin(\theta'))$. One eigenvalue is given by
\footnotesize
\begin{align*}
    \lambda_{0} = \frac{\sqrt{-\cos (2 \theta-2 \theta')-\cos (2 \theta+2 \theta')-2 \cos (2
   \theta)-2 \cos (2 \theta')+6}}{2 \sqrt{2}}
\end{align*} 
\normalsize
and its unnormalized eigenvector $\ket{\tilde{v}_{0}(\theta, \theta')}$ is given by
   \footnotesize
\begin{align*}
    & \Big\{\tan (\theta) \cot (\theta')+\frac{\csc (\theta) \sec (\theta) \csc
   (\theta') \sec (\theta') \left(\Delta-\sin (\theta)+\sin (\theta')\right)
   \delta}{2 \Delta },\\
   & \frac{1}{2} \csc (\theta') \left(\sec (\theta) \Delta -2 \tan
   (\theta)\right),\frac{1}{2} \csc (\theta) \sec (\theta') \left(\Delta -2 \sin
   (\theta')\right),1\Big\},
\end{align*}
\normalsize
where $\Delta \coloneqq \sqrt{-2 \cos (2 \theta) \cos ^2(\theta')-\cos (2 \theta')+3}$ and $\delta = \left(2 \cos (2 \theta) \cos ^2(\theta')+2 \sin (\theta') \Delta +\cos (2
   \theta')-3\right)$. We denote with $\ket{v_{0}(\theta, \theta')} := \frac{\ket{\tilde{v}_{0}(\theta, \theta')}}{\sqrt{\bra{\tilde{v}_{0}(\theta, \theta')}\!\!\ket{\tilde{v}_{0}(\theta, \theta')}}}$ the corresponding normalized vector. 
   
For $U = \px  \otimes \px$, the paired measurement outcome for \begin{align*}
    \hat{\Pi}_{W}(0) = \ket{v_{0}(\theta, \phi)}\!\!\bra{v_{0}(\theta, \phi)} + U\ket{v_{0}(\theta, \phi)}\!\!\bra{v_{0}(\theta, \phi)}U
\end{align*} 
is obtained with probability 1, resulting in a deterministic measurement outcome. 

For simplicity, we abbreviate $\ket{v_{0}(\theta, \theta')}$ as $\ket{v_{0}}$. The post-measurement state is then deterministically given by 
\small
\begin{align*}
    \tilde{\rho}(0, z) & = \begin{pmatrix}
    \bra{v_{0}}\! [W_{\theta} \vnop W_{\theta}](z) \!\ket{v_{0}} & \bra{v_{0}}\! U [W_{\theta} \vnop W_{\theta}](z) \!\ket{v_{0}} \\
    \bra{v_{0}}\! [W_{\theta} \vnop W_{\theta}](z) U \!\ket{v_{0}} & \bra{v_{0}}\! U [W_{\theta} \vnop W_{\theta}](z) U \!\ket{v_{0}}
    \end{pmatrix} \\
    &= \px \begin{pmatrix}
       \frac{1}{4} \left(\Delta+2\right) & \frac{1}{2} \cos (\theta) \cos (\theta') \\
 \frac{1}{2} \cos (\theta) \cos (\theta') & \frac{1}{8} \left(4-2 \Delta \right) \\
    \end{pmatrix} \px.
\end{align*}
\normalsize
Upon applying the unitary $U' = \px \pz$, state $\tilde{\rho}(0, z)$ becomes $\ket{\pm \theta^{\vnop}}\!\!\bra{\pm \theta^{\vnop}}$.
\end{IEEEproof}

\section{Results for the Five-Qubit Factor Graph}

\subsection{Decoding on General BSCQ Channels}

Now, consider a qubit BSCQ $W$ which is neither a pure state channel nor a purely classical channel. Specifically, we consider the channel family defined by
\begin{align} \label{eq:bscpsc}
    W(z) &= (1-p)H\ket{(-1)^{z} \theta}\!\!\bra{(-1)^{z} \theta}H^{\dag} \nonumber \\ & \qquad + p H\ket{(-1)^{z \oplus 1} \theta}\!\!\bra{(-1)^{z \oplus 1} \theta} H^{\dag}
\end{align}
and observe that $W(z) = U^{z} \rho U^{z}$ for $\rho = W(0)$ and $U = \px$.

We note that $W(z)$ may be rewritten in a more general channel formulation $\tilde{W}(z) = (1-q)H\ket{(-1)^{z} \tilde{\theta}}\!\!\bra{(-1)^{z} \tilde{\theta}}H^{\dag} + q \frac{\mathbb{I}}{2}$ by setting 
\begin{align*}
 q &= 1-\sqrt{-2 (p-1) p \cos (2 \theta)+2 (p-1) p+1} \\
 \tilde{\theta} &= \frac{\cot (\theta)-\csc (\theta) \sqrt{-2 (p-1) p \cos (2 \theta)+2 (p-1)
   p+1}}{2 p-1}.
\end{align*}
where for a given $q$, if $\tilde{\theta} = \frac{\pi}{2}$ the corresponding bit flip probability is $\frac{q}{2}$.

PMBPQM enables message-passing decoding for general BSCQ channels.
Here, we plot simulation results for PMBPQM for the root node of the 5-qubit factor graph with parity checks $x_{1} \oplus x_{2} \oplus x_{3} = 0$ and $x_{1} \oplus x_{4} \oplus x_{5} = 0$ depicted in Figure~\ref{fig:fg5}.
We note that one can coherently decode all of the bits by adopting the message-passing framework introduced in~\cite{Piveteau-arxiv21} which introduces a quantum register to store the reliability of the message.

For the five-qubit factor-graph, the PMBPQM approach consists of the following steps:
\begin{enumerate}
\item Implementing paired measurement on qubits 2 and 3 for $W \cnop W$ to get outcome $a$ and post-measurement state $\tilde{\rho}(a, z)$
\item Implementing paired measurement on qubits 4 and 5 for $W \cnop W$ to get outcome $b$ and post-measurement state $\tilde{\rho}(b, z)$
\item Implementing paired measurement for $\tilde{\rho}(a, z) \otimes \tilde{\rho}(b, z)$; obtain measurement outcome $c$ and post-measurement state $\tilde{\rho}(c, z)$
\item Measuring $\tilde{\rho}(c,z)$ and qubit 1 through Helstrom measurement for $\{W(0) \otimes \tilde{\rho}(c, 0), W(1) \otimes \tilde{\rho}(c, 1)\}$
\end{enumerate}

This process is depicted in Figure~\ref{fig:fg5}. We then generate success probabilities for the five-qubit factor-graph using both a paired-measurement BPQM approach and a collective Helstrom measurement respectively. The left hand side of Figure~\ref{fig:hel_pm_reldiff} shows the overall success probability of both approaches, while the right hand side of Figure~\ref{fig:hel_pm_reldiff} plots their relative difference. We observe that the relative difference is small for all choices of channel parameters $(\theta, p)$ and in the special case of a PSC ($p=0$), there is no relative difference.
For the case of $p=0.5$, which is not shown, the channel is worthless the relative difference also vanishes.
This explains why the relative difference first increases with $p$ and then decreases.

We have also considered the ``worst case'' gap between paired-measurement BPQM and the collective Helstrom measurement.
For the tested cases, the worst case gap is relatively small and the paired-measurement BPQM is either optimal or close-to-optimal in all cases. 
We also tested that PMBPQM outperforms or is approximately equivalent to the naive locally-greedy that first measures and the performs classical BP. 
We additionally demonstrate that a small gap between local measurement strategies and the collective Helstrom will also occur with any approach using semi-local binary-output measurements. Namely, there exists an example where no local measurement approach or extension of BPQM can achieve the optimal Helstrom success probability.

\begin{lem}
Consider a tree factor graph for a binary linear code where all bits are sent through CQ channels. When the root bit has value $z\in \{0,1\}$, we denote the density matrix of the observations by $\rho_{z}$. Let $P_{LM}(\rho_{0}, \rho_{1})$ be the optimal success probability for decoding algorithms that start from the bottom of the tree and, in each round, measure a single check node or bit node is conditional on all previous measurement results. Then, there are CQ channels where \vspace*{-1mm}
\begin{align*}
    P_{\text{LM}}(\rho_{0}, \rho_{1}) < P_{\text{H}}\big(\rho_{0}, \rho_{1}, \tfrac{1}{2}\big).
\end{align*}
\vspace{-7mm}
\end{lem}
\vspace{1mm}

\begin{IEEEproof}
Consider the channels $W$ and $W'$ defined by
\begin{align*}
    W(z) &= \px^{z} \begin{pmatrix}
  \frac{2}{3} & \frac{1}{6} \\
  \frac{1}{6} & \frac{1}{3} 
\end{pmatrix}\px^{z} \\
 W'(z) &= \px^{z} \begin{pmatrix}
  \frac{2}{3} & \frac{1}{8} \\
  \frac{1}{8} & \frac{1}{3} 
\end{pmatrix}\px^{z} \textcolor{blue}{.}
\end{align*}
Now we consider an example of a three-qubit (bit node) factor graph where the root qubit and second qubit are both separately sent through $W$ and where the third qubit is sent through $W'$. Then, the Helstrom success probability is given by $\text{P}_{\text{H}} \approx 0.74147$. We now demonstrate that any decoding strategy which involves binary-output measurements on two or fewer qubits will be suboptimal . 

The combined channel for the two children qubits is then $W \vnop W'$; with the local Helstrom measurement for the two children qubits being given by the eigenvectors of $[W \vnop W'](0) - [W \vnop W'](1)$. 

We require that a ``measure up the tree'' approach depend \emph{only} on previous measurement outcomes and the current qubits being measured (rather than on channel parameters of qubits higher up the tree). In the case where all future qubits to be measured are trivially indistinguishable, the problem is effectively reduced to a single-qubit hypothesis testing problem. Thus, each ``measure up the tree'' approach must allow for locally-optimal hypothesis testing.

 This is equivalent to requiring the binary-output measurement to have measurement elements which are equivalent to a grouping of the Helstrom measurement projectors. In this case, the eigenvalues of $[W \vnop W'](0) - [W \vnop W'](1)$ are non-degenerate and so the possible rank-two measurements simply consist of different groupings of the distinct eigenvectors which we denote here by $\Big\{ \ket{\lambda_{1}}\!\!\bra{\lambda_{1}},  \ket{\lambda_{2}}\!\!\bra{\lambda_{2}}, \ket{-\lambda_{1}}\!\!\bra{-\lambda_{1}}, \ket{-\lambda_{2}}\!\!\bra{-\lambda_{2}}\Big\}$.
 We now list the distinct groupings and the corresponding success probability:
\begin{center}
\begin{tabular}{ |c|c|c| } 
\hline
$\Pi_{0}$ & $\mathrm{P}_{\mathrm{succ}}$ \\
\hline
$\ket{\lambda_{1}}\!\!\bra{\lambda_{1}}+   \ket{\lambda_{2}}\!\!\bra{\lambda_{2}}$ & 0.737088 \\ 
$\ket{\lambda_{1}}\!\!\bra{\lambda_{1}}+  \ket{-\lambda_{1}}\!\!\bra{-\lambda_{1}}$ & 0.736276 \\ 
$ \ket{\lambda_{1}}\!\!\bra{\lambda_{1}}+   \ket{-\lambda_{2}}\!\!\bra{-\lambda_{2}}$ & 0.738794 \\ 
\hline
\end{tabular}

\end{center}
It follows that $\text{P}_{\text{LM}}(\rho_{0}, \rho_{1}) \approx 0.738794 < \text{P}_{\text{H}}$. 
\end{IEEEproof}

\section{PMBPQM versus Locally Greedy Algorithms}
\label{PMBPQMLG}

\begin{figure*}[t]
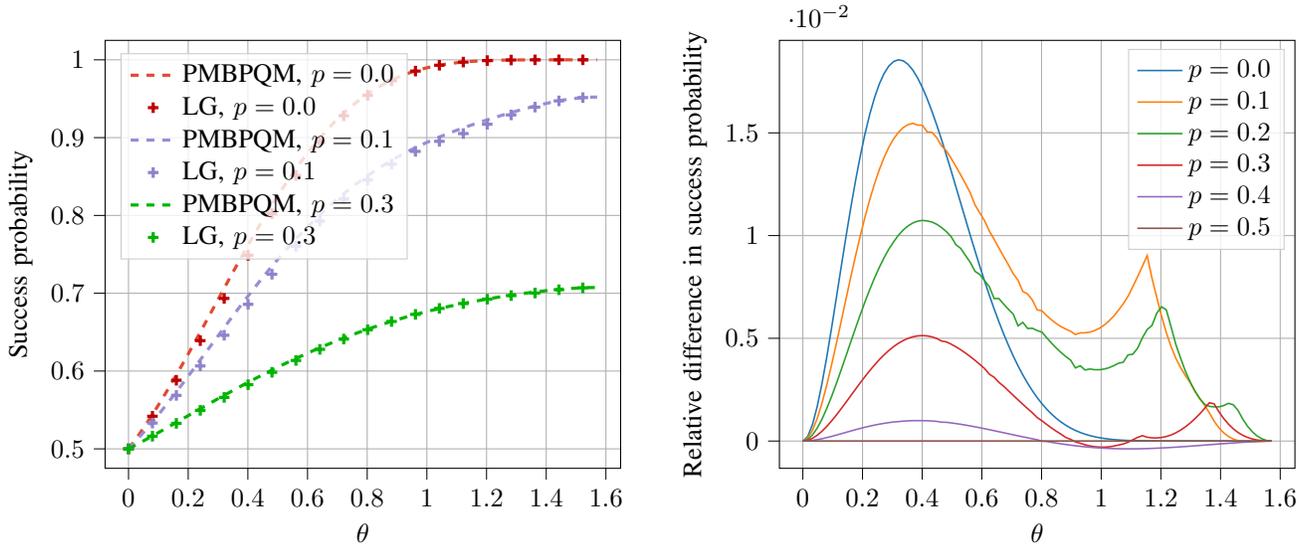

\begin{center}
  \input{LGcomp.tex}
  \hspace{2mm}
  \input{LGrel.tex}
\end{center}
\vspace{-4mm}
  \caption{\label{fig:lg_bpqm}The left hand side depicts success probability for both the locally greedy and PMBPQM approach as a function of $\theta$ for $p\in\{0, 0.1, 0.2\}$. The right hand side depicts the relative difference in success probability between the locally greedy and PMBPQM approach as a function of $\theta$ for $p\in\{0, 0.1, 0.2, 0.3, 0.4\}$. The PMBPQM outperforms or is approximately equivalent to the LG approach for all tested $p$ and $\theta$ combinations.}
\end{figure*}
Finally, we demonstrate that, for the examples tested, PMBPQM outperforms or is approximately equivalent to a naive, locally-greedy measurement approach. A naive locally-greedy decoding approach consists of measuring two qubits (at either a check or bit node) in each round according to the \emph{local} Helstrom measurement for the current prior $q = \text{Pr}(z = 0)$. After each measurement, the prior $q$ is updated using Bayes' Theorem and the next set of two qubits are measured via a local Helstrom measurement. 

We take as an example the 7-qubit factor graph with the following parity checks:
\begin{align*}
    x_{1} &= x_{2} \oplus x_{3}, \ \ x_{1} = x_{4} \oplus x_{5}, \ \ x_{4} = x_{6}, \ \ x_{4} = x_{7}
\end{align*}
Then, assuming that the starting prior is $q = \frac{1}{2}$, a locally greedy approach on the 7-qubit factor graph consists of the following steps:
\begin{enumerate}
    \item Measure qubits 6 and 7 (observations of qubit 4) with the Helstrom measurement for $\{ [W \vnop W](0), [W \vnop W](1) \}$. Update prior probability $q'$ where $q'$ is the probability that the state of qubit 4 is $W(0)$
    \item Measure qubits 4 and 5 with the Helstrom measurement for distinguishing
    \begin{align*}
        q' W(0) \otimes W(0) + (1-q') W(1) \otimes W(1) \\
        q' W(0) \otimes W(1) + (1-q') W(1) \otimes W(0)
    \end{align*}
    Update prior probability $q$
    \item Measure qubits 2 and 3 with the Helstrom measurement for distinguishing $\Big\{ [W \cnop W](0), [W \cnop W](1) \Big\}$ with prior $q$. Update $q$ after measurement
    \item Measure root node with Helstrom for $\{W(0), W(1)\}$ and prior $q$. Update $q$ to find final prior (which is equivalent to the final success probability)
\end{enumerate}
The results of PMBPQM and the locally-greedy approach are depicted in Fig.~\ref{fig:lg_bpqm}.

\section{Density Evolution}

Density evolution is an technique that allows one to analyze the asymptotic performance of a given code ensemble with BP decoding by tracking the probability density function of messages that are passed along edges of a factor graph~\cite{RU-2008}. Density evolution thus determines whether a given channel and degree distribution leads to asymptotically reliable decoding.
In turn, this enables the construction and optimization of low-density parity-check (LDPC) codes that achieve reliable communication at rates close to the channel capacity.

\begin{figure*}[t]
\begin{center}
  \scalebox{0.9}{\input{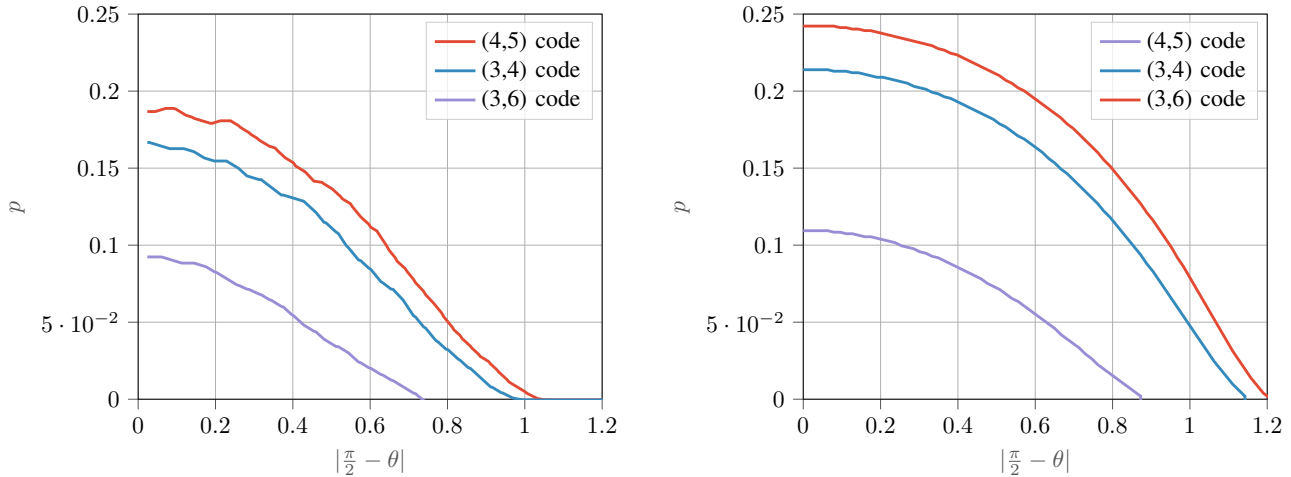}}
  \hspace{2mm}
  \scalebox{0.9}{
\begin{tikzpicture}

\definecolor{darkgray176}{RGB}{176,176,176}
\definecolor{chocolate2267451}{RGB}{226,74,51}
\definecolor{dimgray85}{RGB}{85,85,85}
\definecolor{gainsboro229}{RGB}{229,229,229}
\definecolor{lightgray204}{RGB}{204,204,204}
\definecolor{mediumpurple152142213}{RGB}{152,142,213}
\definecolor{steelblue52138189}{RGB}{52,138,189}

\begin{axis}[
legend cell align={left},
legend style={fill opacity=0.8, draw opacity=1, text opacity=1, draw=lightgray204},
tick align=outside,
tick pos=left,
x grid style={darkgray176},
xlabel=\textcolor{dimgray85}{$|\frac{\pi}{2}-\theta|$},
xmajorgrids,
xmin=0, xmax=1.2,
xtick style={color=dimgray85},
y grid style={darkgray176},
ylabel=\textcolor{dimgray85}{$p$},
ymajorgrids,
ymin=0, ymax=0.25,
ytick style={color=dimgray85}
]

\addplot [very thick, mediumpurple152142213,y filter/.code={\pgfmathparse{\pgfmathresult/2}\pgfmathresult}]
table {%
1.5707963267949 -1
1.55492969723131 -1
1.53906306766773 -1
1.52319643810414 -1
1.50732980854056 -1
1.49146317897697 -1
1.47559654941339 -1
1.4597299198498 -1
1.44386329028622 -1
1.42799666072263 -1
1.41213003115905 -1
1.39626340159546 -1
1.38039677203188 -1
1.36453014246829 -1
1.34866351290471 -1
1.33279688334112 -1
1.31693025377754 -1
1.30106362421395 -1
1.28519699465037 -1
1.26933036508679 -1
1.2534637355232 -1
1.23759710595962 -1
1.22173047639603 -1
1.20586384683245 -1
1.18999721726886 -1
1.17413058770528 -1
1.15826395814169 -1
1.14239732857811 -1
1.12653069901452 -1
1.11066406945094 -1
1.09479743988735 -1
1.07893081032377 -1
1.06306418076018 -1
1.0471975511966 -1
1.03133092163301 -1
1.01546429206943 -1
0.999597662505843 -1
0.983731032942258 -1
0.967864403378674 -1
0.951997773815089 -1
0.936131144251504 -1
0.920264514687919 -1
0.904397885124334 -1
0.88853125556075 -1
0.872664625997165 0.00390625
0.85679799643358 0.009765625
0.840931366869995 0.015625
0.82506473730641 0.021484375
0.809198107742825 0.02734375
0.793331478179241 0.033203125
0.777464848615656 0.0390625
0.761598219052071 0.044921875
0.745731589488486 0.052734375
0.729864959924901 0.05859375
0.713998330361317 0.06640625
0.698131700797732 0.072265625
0.682265071234147 0.078125
0.666398441670562 0.083984375
0.650531812106977 0.091796875
0.634665182543393 0.09765625
0.618798552979808 0.103515625
0.602931923416223 0.109375
0.587065293852638 0.115234375
0.571198664289053 0.12109375
0.555332034725468 0.126953125
0.539465405161884 0.130859375
0.523598775598299 0.13671875
0.507732146034714 0.142578125
0.491865516471129 0.146484375
0.475998886907544 0.150390625
0.460132257343959 0.15625
0.444265627780375 0.16015625
0.42839899821679 0.1640625
0.412532368653205 0.16796875
0.39666573908962 0.171875
0.380799109526035 0.17578125
0.364932479962451 0.1796875
0.349065850398866 0.18359375
0.333199220835281 0.185546875
0.317332591271696 0.189453125
0.301465961708111 0.19140625
0.285599332144526 0.1953125
0.269732702580942 0.197265625
0.253866073017357 0.19921875
0.237999443453772 0.203125
0.222132813890187 0.205078125
0.206266184326602 0.20703125
0.190399554763018 0.208984375
0.174532925199433 0.2109375
0.158666295635848 0.2109375
0.142799666072263 0.212890625
0.126933036508678 0.21484375
0.111066406945094 0.21484375
0.0951997773815088 0.216796875
0.079333147817924 0.216796875
0.0634665182543392 0.21875
0.0475998886907543 0.21875
0.0317332591271695 0.21875
0.0158666295635848 0.21875
0 0.21875
};
\addlegendentry{(4,5) code}
\addplot [very thick, steelblue52138189,y filter/.code={\pgfmathparse{\pgfmathresult/2}\pgfmathresult}]
table {%
1.5707963267949 -1
1.55492969723131 -1
1.53906306766773 -1
1.52319643810414 -1
1.50732980854056 -1
1.49146317897697 -1
1.47559654941339 -1
1.4597299198498 -1
1.44386329028622 -1
1.42799666072263 -1
1.41213003115905 -1
1.39626340159546 -1
1.38039677203188 -1
1.36453014246829 -1
1.34866351290471 -1
1.33279688334112 -1
1.31693025377754 -1
1.30106362421395 -1
1.28519699465037 -1
1.26933036508679 -1
1.2534637355232 -1
1.23759710595962 -1
1.22173047639603 -1
1.20586384683245 -1
1.18999721726886 -1
1.17413058770528 -1
1.15826395814169 -1
1.14239732857811 0.00390625
1.12653069901452 0.01171875
1.11066406945094 0.01953125
1.09479743988735 0.029296875
1.07893081032377 0.0390625
1.06306418076018 0.048828125
1.0471975511966 0.060546875
1.03133092163301 0.072265625
1.01546429206943 0.083984375
0.999597662505843 0.095703125
0.983731032942258 0.107421875
0.967864403378674 0.119140625
0.951997773815089 0.130859375
0.936131144251504 0.142578125
0.920264514687919 0.154296875
0.904397885124334 0.166015625
0.88853125556075 0.17578125
0.872664625997165 0.1875
0.85679799643358 0.197265625
0.840931366869995 0.20703125
0.82506473730641 0.216796875
0.809198107742825 0.2265625
0.793331478179241 0.236328125
0.777464848615656 0.244140625
0.761598219052071 0.25390625
0.745731589488486 0.26171875
0.729864959924901 0.26953125
0.713998330361317 0.27734375
0.698131700797732 0.28515625
0.682265071234147 0.29296875
0.666398441670562 0.298828125
0.650531812106977 0.306640625
0.634665182543393 0.3125
0.618798552979808 0.3203125
0.602931923416223 0.326171875
0.587065293852638 0.33203125
0.571198664289053 0.337890625
0.555332034725468 0.341796875
0.539465405161884 0.34765625
0.523598775598299 0.353515625
0.507732146034714 0.357421875
0.491865516471129 0.36328125
0.475998886907544 0.3671875
0.460132257343959 0.37109375
0.444265627780375 0.375
0.42839899821679 0.37890625
0.412532368653205 0.3828125
0.39666573908962 0.38671875
0.380799109526035 0.390625
0.364932479962451 0.392578125
0.349065850398866 0.396484375
0.333199220835281 0.3984375
0.317332591271696 0.40234375
0.301465961708111 0.404296875
0.285599332144526 0.40625
0.269732702580942 0.41015625
0.253866073017357 0.412109375
0.237999443453772 0.4140625
0.222132813890187 0.416015625
0.206266184326602 0.41796875
0.190399554763018 0.41796875
0.174532925199433 0.419921875
0.158666295635848 0.421875
0.142799666072263 0.423828125
0.126933036508678 0.423828125
0.111066406945094 0.42578125
0.0951997773815088 0.42578125
0.079333147817924 0.42578125
0.0634665182543392 0.427734375
0.0475998886907543 0.427734375
0.0317332591271695 0.427734375
0.0158666295635848 0.427734375
0 0.427734375
};
\addlegendentry{(3,4) code}

\addplot [very thick, chocolate2267451,y filter/.code={\pgfmathparse{\pgfmathresult/2}\pgfmathresult}]
table {%
1.5707963267949 -1
1.55492969723131 -1
1.53906306766773 -1
1.52319643810414 -1
1.50732980854056 -1
1.49146317897697 -1
1.47559654941339 -1
1.4597299198498 -1
1.44386329028622 -1
1.42799666072263 -1
1.41213003115905 -1
1.39626340159546 -1
1.38039677203188 -1
1.36453014246829 -1
1.34866351290471 -1
1.33279688334112 -1
1.31693025377754 -1
1.30106362421395 -1
1.28519699465037 -1
1.26933036508679 -1
1.2534637355232 -1
1.23759710595962 -1
1.22173047639603 -1
1.20586384683245 0.001953125
1.18999721726886 0.0078125
1.17413058770528 0.017578125
1.15826395814169 0.02734375
1.14239732857811 0.0390625
1.12653069901452 0.05078125
1.11066406945094 0.0625
1.09479743988735 0.076171875
1.07893081032377 0.08984375
1.06306418076018 0.103515625
1.0471975511966 0.1171875
1.03133092163301 0.130859375
1.01546429206943 0.14453125
0.999597662505843 0.158203125
0.983731032942258 0.171875
0.967864403378674 0.18359375
0.951997773815089 0.197265625
0.936131144251504 0.208984375
0.920264514687919 0.220703125
0.904397885124334 0.232421875
0.88853125556075 0.2421875
0.872664625997165 0.25390625
0.85679799643358 0.263671875
0.840931366869995 0.2734375
0.82506473730641 0.283203125
0.809198107742825 0.29296875
0.793331478179241 0.302734375
0.777464848615656 0.310546875
0.761598219052071 0.3203125
0.745731589488486 0.328125
0.729864959924901 0.3359375
0.713998330361317 0.34375
0.698131700797732 0.3515625
0.682265071234147 0.357421875
0.666398441670562 0.365234375
0.650531812106977 0.37109375
0.634665182543393 0.376953125
0.618798552979808 0.3828125
0.602931923416223 0.388671875
0.587065293852638 0.39453125
0.571198664289053 0.400390625
0.555332034725468 0.404296875
0.539465405161884 0.41015625
0.523598775598299 0.4140625
0.507732146034714 0.419921875
0.491865516471129 0.423828125
0.475998886907544 0.427734375
0.460132257343959 0.431640625
0.444265627780375 0.435546875
0.42839899821679 0.439453125
0.412532368653205 0.443359375
0.39666573908962 0.447265625
0.380799109526035 0.44921875
0.364932479962451 0.453125
0.349065850398866 0.455078125
0.333199220835281 0.458984375
0.317332591271696 0.4609375
0.301465961708111 0.462890625
0.285599332144526 0.46484375
0.269732702580942 0.466796875
0.253866073017357 0.46875
0.237999443453772 0.470703125
0.222132813890187 0.47265625
0.206266184326602 0.474609375
0.190399554763018 0.4765625
0.174532925199433 0.478515625
0.158666295635848 0.478515625
0.142799666072263 0.48046875
0.126933036508678 0.48046875
0.111066406945094 0.482421875
0.0951997773815088 0.482421875
0.079333147817924 0.484375
0.0634665182543392 0.484375
0.0475998886907543 0.484375
0.0317332591271695 0.484375
0.0158666295635848 0.484375
0 0.484375
};
\addlegendentry{(3,6) code}
\end{axis}

\end{tikzpicture}}
  \vspace{0mm}
\end{center}
  \caption{\label{fig:de}The left diagram shows the noise threshold boundary for LDPC codes on~\eqref{eq:bscpsc} as a function of $|\frac{\pi}{2} - \theta|$ and $p=q/2$ for $W_{\theta, q}(z) = (1-q)\sketbra{(-1)^{z} \theta} + q \frac{\mathbb{I}}{2}$, where $(\theta,q)=(0,2p)$ is a BSC with bit flip probability~$p$. Results are provided for the $(d_v,d_c)$ regular LDPC codes indicated in the legend. On the right, we see a corresponding diagram for the Holevo bounds of the associated code rates, $\{0.2, 0.25, 0.5\}$.}  \vspace*{-3mm}
\end{figure*}

While there is no currently known method to analyze the performance of general protocols such as the collective Helstrom measurement, we demonstrate here that it is possible to extend density evolution to provide an asymptotic analysis (via Monte Carlo simulation) of PMBPQM. Such a method is possible because each the paired measurement operation allows the result of bit- and check-node combining to be viewed as a new single-qubit BSCQ channel.
Thus, the paired measurements can be iteratively performed up the tree and the threshold can be found by tracking the channel parameters after a large number of channel combining iterations. 

We begin by describing bit- and check-node combining operations for higher-degree bit and check nodes. First, we consider the case where the bit node has degree $d_{v}$ and the check node has degree $d_{c}$, and where for each node the channel corresponding to the $j^{th}$ qubit is denoted as $W^{(j)}$. The steps are then given by:
\begin{enumerate}
    \item For $j \!\in\! \{1,\ldots, \lfloor \frac{d_{v}}{2} \rfloor\}$, execute paired measurement $\big\{ [W^{(2j-1)} \!\!\vnop\!\! W^{(2j)}](0),[W^{(2j-1)} \!\!\vnop\!\! W^{(2j)}](1)\big\}$ to get post-measurement channels $\{ W^{(12)}, \ldots, W^{(d_{v}-1, d_v)}\}$ (or $W^{(d_{v}-1)}$ if $d_{v}$ is odd)
    \item Repeat the above step with the post-measurement channel set to form the new set $\{W^{(1234)}$, $W^{(5678)}, \ldots\}$, etc.
    \item Repeat until only a single post-measurement qubit remains with effective channel $\tilde{W}$
    \item Implement a final paired measurement for $\tilde{W} \vnop W$
\end{enumerate}
\vspace{1mm}
For a check node, the steps are analogous (with bit-node combining replaced by check-node combining) except there is no additional combining with $W$ (in the 4th step). For details of general qubit BSCQ channel combining at bit and check nodes, see Appendix~\ref{BSCQgen}.

Density evolution works by tracking the distribution of the PMBPQM qubit channels.
We implement $N$ stages this using the Monte Carlo process described below.
For a regular LDPC code with bit and check degrees $d_v$ and $d_c$, the process alternates between bit and check channel combining.  To start, we assume that all bits are initially sent separately through the BSCQ channel $W_{\theta, p}(z) = (1-p)\sketbra{(-1)^{z} \theta} + p \frac{\mathbb{I}}{2}$.

The following method is used for density evolution: 
\begin{enumerate}
    \item Create multiset $\mathcal{S} = \{(\theta, p)\}|_{k=1}^{M}$  with $M=5000$ copies of the channel parameters and initialize multiset $\mathcal{S}' = \emptyset$
    \item For $j$ in $\{1, 2,\ldots,N\}$:
        \begin{enumerate}
            \item Randomly draw $(\theta, q)$ and $(\theta', q')$ from $\mathcal{S}$
            \item If $j$ is odd, implement iterative paired measurements for the check node until all $d_{c}$ qubits are condensed to a single post-measurement qubit state $\tilde{\rho}(j, z) = (1-q_{j}) \sketbra{(-1)^{z}\theta_{j}} + q_{j} \frac{\mathbb{I}}{2}$ 
            \item If $j$ is even, implement iterative paired measurements for the check node until all $d_{v}$ qubits and the parent qubit (with parameters $(\theta, q)$) are condensed to a single post-measurement qubit state,  $\tilde{\rho}(j, z) = (1-q_{j}) \sketbra{(-1)^{z}\theta_{j}} + q_{j} \frac{\mathbb{I}}{2}$
            \item Add the new channel parameters $(\theta_{j}, q_{j})$ to $\mathcal{S}'$
            \item Repeat 2a-2d until $\mathcal{S}'$ has $M$ elements, set $\mathcal{S} = \mathcal{S}'$
        \end{enumerate}
  \item The overall success probability is the average over $(\theta,q)\in \mathcal{S}$ of the success probability for distinguishing between $(1-q) \sketbra{\theta} + q \frac{\mathbb{I}}{2}$ and $(1-q) \sketbra{-\theta} + q \frac{\mathbb{I}}{2}$
\end{enumerate}

We say the channel $W_{\theta,p}$ is below the noise threshold if the success probability for the root bit approaches 1 as $N \rightarrow \infty$.
The noise thresholds and Holevo limits are shown in Figure~\ref{fig:de}.

\section{Conclusions}

We introduce a paired-measurement BPQM protocol that extends BPQM to the case of BSCQ channels.
This protocol reduces to BP for classical symmetric channels and matches the known (optimal) BPQM protocol for the pure-state channel. For more general BSCQ channels, we show that PMBPQM performs well in all tested instances. We also consider the application of PMBPQM to large factor graphs by using asymptotic analysis techniques. Based on Monte Carlo density evolution, we plot the region of channel parameters for which PMBPQM achieves asymptotically reliable decoding for a few different regular LDPC codes on BSCQ channels. The introduction of PMBPQM naturally leads to interesting open questions about the ``worst case'' gap between paired-measurement BPQM and the collective Helstrom measurement; as well as whether there exists an extension of BPQM that is optimal for general BSCQ channels.  

\section{Acknowledgments}
The authors would like to thank Narayanan Rengaswamy, Joseph Renes, and Saikat Guha for helpful discussions during this research.
The presentation of material was also improved significantly by comments of anonymous reviewers.
This work was supported in part by the National Science Foundation (NSF) under Grants No. 1908730 and 1910571. Any opinions, findings, conclusions, and recommendations expressed in this material are those of the authors and do not necessarily reflect the views of these sponsors.

\newpage

\IEEEtriggeratref{11}

\begin{thebibliography}{10}

\bibitem{Gallager-60}
R.~G. Gallager, {\em Low-Density Parity-Check Codes}.
\newblock PhD thesis, M.I.T., Cambridge, MA, USA, 1960.

\bibitem{Berrou-icc93}
C.~Berrou, A.~Glavieux, and P.~Thitimajshima, ``Near {Shannon} limit
  error-correcting coding and decoding: Turbo-codes,'' in {\em Proc.\ IEEE
  Int.\ Conf.\ Commun.}, vol.~2, (Geneva, Switzerland), pp.~1064--1070, IEEE,
  May 1993.

\bibitem{Mackay-it99}
D.~J.~C. MacKay, ``Good error-correcting codes based on very sparse matrices,''
  {\em IEEE Trans.\ Inform.\ Theory}, vol.~45, pp.~399--431, March 1999.

\bibitem{Pearl-aaai82}
J.~Pearl, ``Reverend {B}ayes on inference engines: A distributed hierarchical
  approach,'' in {\em AAAI Conf.\ Artificial Intelligence}, 1982.

\bibitem{McEliece-jsac98}
R.~J. McEliece, D.~J.~C. MacKay, and J.~Cheng, ``Turbo decoding as an instance
  of {P}earl's "belief propagation" algorithm,'' {\em IEEE J.\ Select.\ Areas
  Commun.}, vol.~16, pp.~140--152, Feb. 1998.

\bibitem{Kschischang-jsac98}
F.~R. Kschischang and B.~J. Frey, ``Iterative decoding of compound codes by
  probability propagation in graphical models,'' {\em IEEE J.\ Select.\ Areas
  Commun.}, vol.~16, no.~2, pp.~219--230, 1998.

\bibitem{PhysRevA.56.131}
B.~Schumacher and M.~D. Westmoreland, ``Sending classical information via noisy
  quantum channels,'' {\em Phys. Rev. A}, vol.~56, pp.~131--138, Jul 1997.

\bibitem{Wilde2013}
M.~M. Wilde and S.~Guha, ``Polar codes for classical-quantum channels,'' {\em
  IEEE Transactions on Information Theory}, vol.~59, pp.~1175--1187, Feb 2013.

\bibitem{Lloyd2012}
V.~Giovannetti, S.~Lloyd, and L.~Maccone, ``Achieving the {H}olevo bound via
  sequential measurements,'' {\em Physical Review A}, vol.~85, Jan 2012.

\bibitem{PhysRevA.105.032446}
A.~Jagannathan, M.~Grace, O.~Brasher, J.~H. Shapiro, S.~Guha, and J.~L. Habif,
  ``Demonstration of quantum-limited discrimination of multicopy pure versus
  mixed states,'' {\em Phys. Rev. A}, vol.~105, p.~032446, Mar 2022.

\bibitem{PhysRevLett.118.040801}
Q.~Zhuang, Z.~Zhang, and J.~H. Shapiro, ``Optimum mixed-state discrimination
  for noisy entanglement-enhanced sensing,'' {\em Phys. Rev. Lett.}, vol.~118,
  p.~040801, Jan 2017.

\bibitem{da_Silva_2013}
M.~P. da~Silva, S.~Guha, and Z.~Dutton, ``Achieving minimum-error
  discrimination of an arbitrary set of laser-light pulses,'' {\em Physical
  Review A}, vol.~87, may 2013.

\bibitem{PhysRevLett.106.240502}
S.~Guha, ``Structured optical receivers to attain superadditive capacity and
  the {H}olevo limit,'' {\em Phys. Rev. Lett.}, vol.~106, p.~240502, Jun 2011.

\bibitem{Renes-njp17}
J.~M. Renes, ``Belief propagation decoding of quantum channels by passing
  quantum messages,'' {\em New Journal of Physics}, vol.~19, no.~7, p.~072001,
  2017.

\bibitem{Rengaswamy-isit20b}
N.~Rengaswamy, K.~P. Seshadreesan, S.~Guha, and H.~D. Pfister, ``Quantum
  advantage via qubit belief propagation,'' in {\em Proc.\ IEEE Int.\ Symp.\
  Inform.\ Theory}, pp.~1824--1829, 2020.

\bibitem{Rengaswamy-npjqi21}
N.~Rengaswamy, K.~P. Seshadreesan, S.~Guha, and H.~D. Pfister, ``Belief
  propagation with quantum messages for quantum-enhanced classical
  communications,'' {\em npj Quantum Information}, vol.~7, no.~1, pp.~1--12,
  2021.

\bibitem{Piveteau-arxiv21}
C.~Piveteau and J.~M. Renes, ``Quantum message-passing algorithm for optimal
  and efficient decoding,'' {\em arXiv preprint arXiv:2109.08170}, 2021.

\bibitem{RU-2008}
T.~J. Richardson and R.~L. Urbanke, {\em Modern Coding Theory}.
\newblock New York, NY: Cambridge University Press, 2008.

\bibitem{Richardson-it01}
T.~J. Richardson and R.~L. Urbanke, ``The capacity of low-density parity-check
  codes under message-passing decoding,'' {\em IEEE Trans.\ Inform.\ Theory},
  vol.~47, pp.~599--618, Feb. 2001.

\bibitem{Helstrom-jsp69}
C.~W. Helstrom, ``Quantum detection and estimation theory,'' {\em Journal of
  Statistical Physics}, vol.~1, no.~2, pp.~231--252, 1969.

\bibitem{Renes-it18}
J.~M. Renes, ``Duality of channels and codes,'' {\em IEEE Trans.\ Inform.\
  Theory}, vol.~64, no.~1, pp.~577--592, 2018.

\end{thebibliography}

\vfill

\pagebreak
\begin{appendices}
\onecolumn
\section{Paired Measurement Combining for Qubit BSCQ Channels}
\label{BSCQgen}
Recall that any qubit density matrix is positive semidefinite with unit trace and, thus, can be written as
\[ \rho(\delta, \gamma) \coloneqq \begin{pmatrix} \delta & \gamma \\ \gamma^{*} & 1-\delta
\end{pmatrix}, \]
where $\delta \in [0,1]$ and $\gamma \in \mathbb{C}$ satisfies $4\delta(1-\delta)-4|\gamma|^2 \in [0,1]$.
Such matrices can also be expressed equivalently in the $(\theta,q)$ domain via $(1-q)\sketbra{\theta} + q \frac{\mathbb{I}}{2}$. 
We prefer the former notation in this section and give formulas for the updated post-measurement state after channel combining.

\subsection{Bit-Node Combining}

Consider the bit node combining operation for $W_{\theta, q} \vnop W_{\theta', q'}$. This corresponds to implementing a paired measurement on $\Big\{ \px^{z} \rho(\delta_{1}, \gamma_{1})\px^{z} \otimes \px^{z} \rho(\delta_{2}, \gamma_{2}) \px^{z} \Big\}|_{z=0}^{1}$ where converting the parameters from $(\theta,q)$ to $(\delta,\gamma)$ gives 
\begin{align*}
  \delta_{1} &= \frac{q}{2} + \frac{1}{2} (1 - q) (1 - \sin(\theta)), \ \ \delta_{2} = \frac{q'}{2} + \frac{1}{2} (1 - q') (1 - \sin(\theta')) \\
  \gamma_{1} &= \frac{1}{2} (1 - q) \cos(\theta), \ \ \gamma_{2} = \frac{1}{2} (1 - q') \cos(\theta').
\end{align*}
The paired measurement depends on the eigenvalues and eigenvectors of
\begin{align*}
M(\delta_{1}, \delta_{2}, \gamma_{1}, \gamma_{2}) &= \rho(\delta_{1}, \gamma_{1}) \otimes \rho(\delta_{2}, \gamma_{2}) - \px \rho(\delta_{1}, \gamma_{1})\px \otimes \px \rho(\delta_{2}, \gamma_{2}) \px.
\end{align*}

The expression for the general case where $\gamma_{1} \ne \gamma_{2}$ is lengthy. For the case where $\gamma_{1}=\gamma_{2}$ and $\delta_{1} = \delta_{2}$, we have
\small
\begin{align*}
\tilde{\rho}(0, \delta, \gamma) &= \begin{pmatrix}
 \frac{1}{2} & -2 \gamma^2 \\
 -2 \gamma^2 & \frac{1}{2} \\
    \end{pmatrix} \\
    \tilde{\rho}(1, \delta, \gamma) &= \begin{pmatrix}
    \frac{2\delta^2-2\delta(4\sqrt{4\gamma^2+1}\gamma^2+\sqrt{4\gamma^2+1}+1)+\gamma^2(4\sqrt{4\gamma^2+1}+6)+\sqrt{4\gamma^2+1}+1}{4(\delta-1)\delta+12\gamma^2+2} & \frac{2\gamma^2(-2(\delta-1)\delta+2\gamma^2+1)}{2(\delta-1)\delta+6\gamma^2+1} \\
    \frac{2\gamma^2(-2(\delta-1)\delta+2\gamma^2+1)}{2(\delta-1)\delta+6\gamma^2+1} & \frac{2\delta^2+2\delta(4\sqrt{4\gamma^2+1}\gamma^2+\sqrt{4\gamma^2+1}-1)+\gamma^2(6-4\sqrt{4\gamma^2+1})-\sqrt{4\gamma^2+1}+1}{4(\delta-1)\delta+12\gamma^2+2}
    \end{pmatrix}, \\ 
\end{align*}
\normalsize
where $\tilde{\rho}(0, \delta, \gamma)$ is the post-measurement state given that the measurement outcome is $\frac{\ket{v_{0}}\!\!\bra{v_{0}}}{\bra{v_{0}}\!\!\ket{v_{0}}} + \px \frac{\ket{v_{0}}\!\!\bra{v_{0}}}{\bra{v_{0}}\!\!\ket{v_{0}}}  \px$ and $\tilde{\rho}(1, \delta, \gamma)$ is the post-measurement state given that the measurement outcome is $\frac{\ket{v_{1}}\!\!\bra{v_{1}}}{\bra{v_{1}}\!\!\ket{v_{1}}} + \px \frac{\ket{v_{1}}\!\!\bra{v_{1}}}{\bra{v_{1}}\!\!\ket{v_{1}}}  \px$.
Numerical functions for calculating the updated channel parameters, for the general case where $\gamma_{1} \ne \gamma_{2}$ or $\delta_{1} \ne \delta_{2}$, are provided in the GitHub repository \url{https://github.com/SarahBrandsen/BPQM}.
The corresponding (unnormalized) eigenvectors are
\begin{align*}
\ket{v_{0}} &= \left\{1,-\frac{\sqrt{\frac{1}{\gamma_{1}^2}+4} \gamma_{1}+1}{2
   \gamma_{1}},\frac{\sqrt{\frac{1}{\gamma_{1}^2}+4} \gamma_{1}-1}{2
   \gamma_{1}},1\right\} \\
\ket{v_{1}} &= \left\{-\frac{2 \gamma_{1}^2+\sqrt{4 \gamma_{1}^2+1}+1}{2
   \gamma_{1}^2},-\frac{\sqrt{4 \gamma_{1}^2+1}+1}{2 \gamma_{1}},-\frac{\sqrt{4
   \gamma_{1}^2+1}+1}{2 \gamma_{1}},1\right\}.
\end{align*}
It follows that 
\begin{align*}
    \tilde{\delta}(0,\delta, \gamma) &= \frac{1}{2} \\
    \tilde{\gamma}(0, \delta, \gamma) &= -2\gamma^2 \\
    \tilde{\delta}(1, \delta, \gamma) &= 
    \frac{2\delta^2-2\delta(4\sqrt{4\gamma^2+1}\gamma^2+\sqrt{4\gamma^2+1}+1)+\gamma^2(4\sqrt{4\gamma^2+1}+6)+\sqrt{4\gamma^2+1}+1}{4(\delta-1)\delta+12\gamma^2+2} \\ 
    \tilde{\gamma}(1, \delta, \gamma) &=  \frac{2\gamma^2(-2(\delta-1)\delta+2\gamma^2+1)}{2(\delta-1)\delta+6\gamma^2+1},
\end{align*}
where the probability of obtaining measurement outcome $0$ is 
\begin{align*}
    p_{0} &= \frac{2 (\delta_{1}-1) \delta_{1}+6 \gamma_{1}^2+1}{ 4 \gamma_{1}^2+1}.
\end{align*}

\subsection{Check-Node Combining}

Consider a check node combining operation for $W_{\theta, q} \cnop W_{\theta', q'}$. The corresponding paired measurement is given by the eigendecomposition of 
\begin{align*}
 M \coloneqq \Big( \rho(\delta_{1}, \gamma_{1}) \otimes \rho(\delta_{2}, \gamma_{2}) + \sigma_{x} \rho(\delta_{1}, \gamma_{1}) \sigma_{x} \otimes \sigma_{x} \rho(\delta_{2}, \gamma_{2}) \sigma_{x} \Big) - \Big( \rho(\delta_{1}, \gamma_{1}) \otimes \sigma_{x} \rho(\delta_{2}, \gamma_{2})\sigma_{x} + \sigma_{x} \rho(\delta_{1}, \gamma_{1}) \sigma_{x} \otimes  \rho(\delta_{2}, \gamma_{2}) \Big), 
\end{align*}
where, like the bit node case, converting the parameters from $(\theta,q)$ to $(\delta,\gamma)$ gives 
\begin{align*}
  \delta_{1} &= \frac{q}{2} + \frac{1}{2} (1 - q) (1 - \sin(\theta)), \ \ \delta_{2} = \frac{q'}{2} + \frac{1}{2} (1 - q') (1 - \sin(\theta')) \\
  \gamma_{1} &= \frac{1}{2} (1 - q) \cos(\theta), \ \ \gamma_{2} = \frac{1}{2} (1 - q') \cos(\theta').
\end{align*}

In this case, we have eigenvalue degeneracy in the positive eigenspace of $M$ spanned by vectors $\ket{00}$ and $\ket{11}$. To resolve the degenracy, we choose $\alpha$ to maximize the mutual information of the postmeasurement state with the remainder of the factor graph, which we denote as $\rho_{r}(z)$.
The resulting eigenvectors which generate the paired-measurement are
\begin{align*}
    \ket{v_{0}(\alpha)} &= \frac{1}{\sqrt{1+ \alpha^{2}}}\{1, 0, 0, \alpha\} \\
\ket{v_{1}(\alpha)} &= \frac{1}{\sqrt{1+ \alpha^{2}}}\{-1, 0, 0, \alpha\}.
\end{align*}

The joint post-measurement state (including the remainder of the factor graph) given that we generate the paired measurement based on vectors $\ket{v_{0}}$ and $\ket{v_{1}}$ may then be written as
\begin{align*}
   \omega(\alpha, z) &=  \rho_{r}(z) \otimes \px^{z} \Bigg( \begin{pmatrix}
    \frac{1}{2} \left(\frac{4 \alpha  \gamma_{1} \gamma_{2}}{\alpha ^2+1}+ 2
   \delta_{1} \delta_{2}-\delta_{1}-\delta_{2}+1\right) & \frac{\alpha  \gamma_{2}}{\alpha ^2+1}+\frac{\gamma_{1}}{2} \\
 \frac{\alpha  \gamma_{2}}{\alpha ^2+1}+\frac{\gamma_{1}}{2} & \frac{1}{2} \left(\frac{4 \alpha 
   \gamma_{1} \gamma_{2}}{\alpha ^2+1}-2 \delta_{1} \delta_{2}+\delta_{1}+\delta_{2}\right) \\
   \end{pmatrix} \px^{z} \otimes \ket{0}\!\!\bra{0} \\
   & +  \px^{z} \begin{pmatrix}
   \frac{1}{2} \left(-\frac{4 \alpha  \gamma_{1} \gamma_{2}}{\alpha ^2+1}+2 \delta_{1}
   \delta_{2}-\delta_{1}-\delta_{2}+1\right) & \frac{\gamma_{1}}{2}-\frac{\alpha  \gamma_{2}}{\alpha ^2+1} \\
 \frac{\gamma_{1}}{2}-\frac{\alpha  \gamma_{2}}{\alpha ^2+1} & \frac{1}{2} \left(-\frac{4 \alpha 
   \gamma_{1} \gamma_{2}}{\alpha ^2+1}-2 \delta_{1} \delta_{2}+\delta_{1}+\delta_{2}\right) \\
   \end{pmatrix} \px^{z} \otimes \ket{1}\!\!\bra{1}
   \Bigg).
\end{align*}
\normalsize

The mutual information between the quantum system and the classical observation $z$ is then given by
\begin{align*}
    I_{\alpha}(\rho_{\text{tot}} : Z) & \coloneqq S(Z) + S\Big( \sum_{z=0}^{1} \text{Pr}(Z=z) \times \omega(\alpha, z) \Big) - S \Big( \sum_{z=0}^{1} \text{Pr}(Z=z) \times \omega(\alpha, z) \otimes \ket{z}\!\!\bra{z} \Big).
\end{align*}
In a large number of numerical tests, $\alpha = 1$ always maximizes $I_{\alpha}(\rho_{\text{tot}} : Z)$, so we choose $\alpha = 1$. This leads to post-measurement parameters given by
\begin{align*}
 \tilde{\delta}(0, \gamma_{1}, \gamma_{2}, \delta_{1}, \delta_{2}) &= \frac{1}{2} \left(2  \gamma_{1} \gamma_{2} + 2
   \delta_{1} \delta_{2}-\delta_{1}-\delta_{2}+1\right) \\ \tilde{\gamma}(0, \gamma_{1}, \gamma_{2}, \delta_{1}, \delta_{2}) &= \frac{1}{2}(\gamma_{1} + \gamma_{2}) \\
    \tilde{\delta}(1, \gamma_{1}, \gamma_{2}, \delta_{1}, \delta_{2}) &= \frac{1}{2} \left(- 2  \gamma_{1} \gamma_{2}+2 \delta_{1}
   \delta_{2}-\delta_{1}-\delta_{2}+1\right) \\
    \tilde{\gamma}(1, \gamma_{1}, \gamma_{2}, \delta_{1}, \delta_{2}) &= \frac{1}{2} (\gamma_{1} - \gamma_{2})
\end{align*}
and $p_{0} = \frac{1}{2} + 2 \gamma_{1} \gamma_{2}$ is the probability of obtaining measurement outcome $0$.
\end{appendices}

\end{document}